\documentclass[12pt,preprint]{aastex}

\shorttitle{Planet Signatures}
\shortauthors{Kuchner \& Holman}

\begin{document}

\slugcomment{Revised for ApJ February 27, 2003}

\title{The Geometry of Resonant Signatures in Debris Disks \\ with Planets}

\author{Marc J. Kuchner\altaffilmark{1} and Matthew J. Holman}

\email{mkuchner@cfa.harvard.edu, mholman@cfa.harvard.edu}

\affil{Harvard-Smithsonian Center for Astrophysics, 
60 Garden Street, Cambridge, MA 02138}

\altaffiltext{1}{Michelson Postdoctoral Fellow}

\begin{abstract}

Using simple geometrical arguments,
we paint an overview of the variety of resonant
structures a single planet with moderate eccentricity ($e \lesssim 0.6$)
can create in a dynamically cold, optically thin dust disk.  
This overview may serve as a key for interpreting images of
perturbed debris disks and inferring the dynamical properties of
the planets responsible for the perturbations.  
We compare the resonant geometries found in the solar system dust cloud
with observations of dust clouds around Vega, $\epsilon$ Eridani and
Fomalhaut.
\end{abstract}


\section{Introduction}

Direct imaging of nearby stars can not yet detect light from
extrasolar planets.  However, imaging can detect
circumstellar dust, and when a planet orbits inside a dust cloud, the planet can
reshape the cloud dynamically, as the Earth perturbs
the solar dust cloud.  Several debris disks around nearby main sequence
stars show structures and asymmetries which have been ascribed to
planetary perturbations \citep{burr95, holl98, schn99, koer01, fomalhaut};
perhaps these perturbed disks are signposts of extrasolar planetary systems.

Many of these disk features can be modeled as dust trapped in mean
motion resonances (MMRs) with a planet.
\citet{gold75} suggested that as interplanetary dust spirals into the sun under
the influence of Poynting-Robertson drag (P-R drag), planets could
temporarily trap the dust in MMRs, creating
ring-like density enhancements in the interplanetary cloud.  Since
then, both the InfraRed Astronomical Satellite (IRAS) and the
Diffuse InfraRed Background Explorer (DIRBE) on the Cosmic Background
Explorer (COBE) satellite have provided evidence for a ring of dust particles
trapped by the Earth \citep{jack89,reac91,marz94,derm94,reac95}.
Models of Kuiper Belt dust dynamics \citep{liou99} suggest
that Neptune may also trap dust in first-order MMRs.

Other stars may host planets like the Earth or Neptune.
However, most of the known extrasolar planets do not
resemble the Earth or Neptune; they have masses in the range of 0.3--15
Jupiter masses, and they often have significant
orbital eccentricities  (see, e.g., the review by Marcy \& Butler 2000).
Simulations by \citet{kuch01} show that planets as massive as these
on eccentric orbits placed in a cloud of inspiraling dust often create
two concentrations of dust placed asymmetrically with respect to the star.  
Maps of the vicinity of Vega made with the IRAM
Plateau de Bure interferometer at 1.3~mm \citep{wiln02}
and with the JCMT at 850~$\mu$m \citep{holl98} reveal two concentrations of
circumstellar emission whose asymmetries can be naturally
explained by such a model, possibly indicating the presence
of a few-Jupiter mass planet in an eccentric orbit around
Vega \citep{wiln02}.  Other papers have numerically explored
the interactions of particular planetary system configurations with
a dust disk, with a view towards developing a general key for interpreting
disk structures \citep{roqu94, leca96, liou99, quil02}.

We assemble a primitive version of such a key by mapping the geometries
of the MMRs which are likely to trap the most dust near a planet embedded
in a debris disk.  We illustrate the patterns formed by the libration centers
of the trapped particles in an inertial frame---the frame of a distant observer.
These basic patterns allow us to characterize four structures
that probably span the range of high-contrast
resonant structures a planet on an orbit with eccentricity up to
$\sim 0.6$ and low inclination can create in a dust disk.

\begin{figure}
\epsscale{0.9}
\plotone{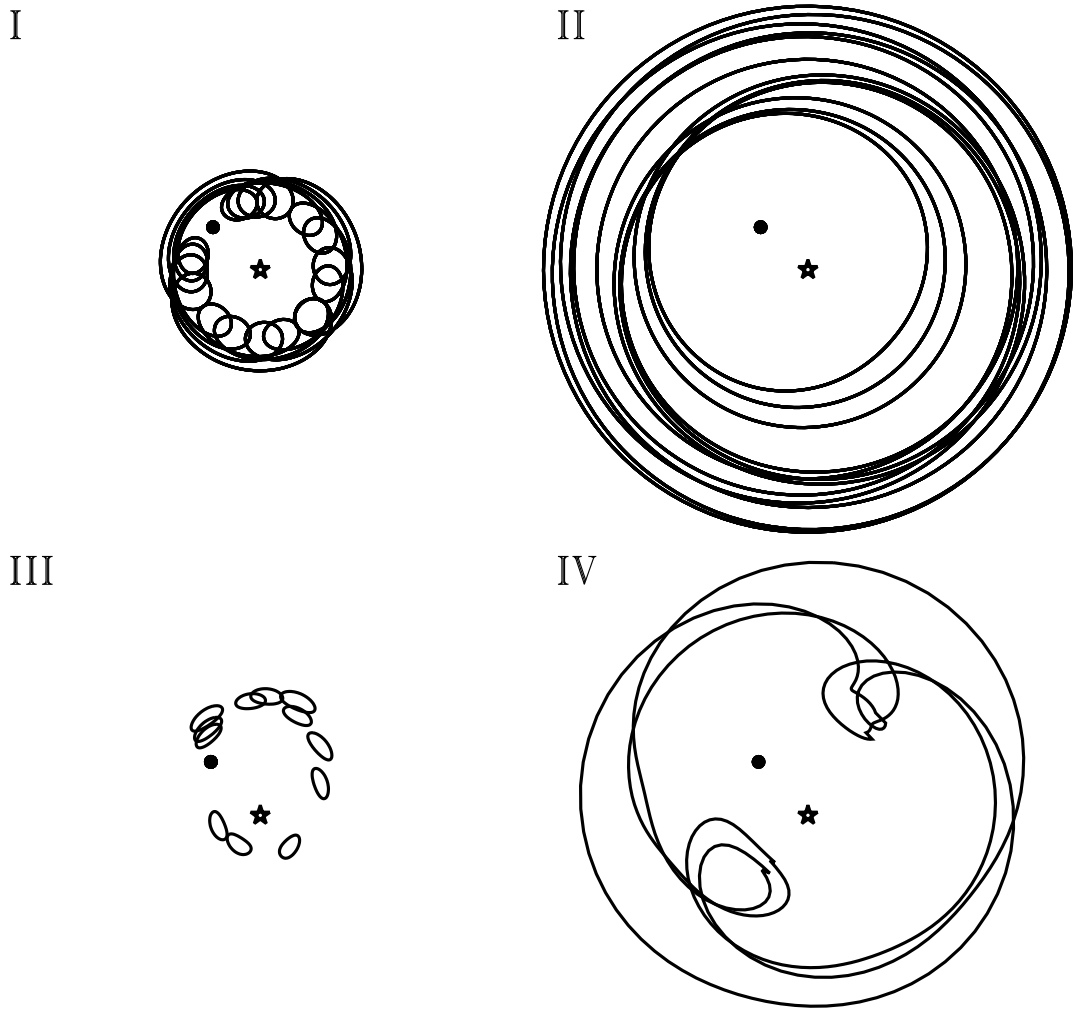}
\figcaption{Four basic resonant structures: 
I) low mass planet on a low eccentricity orbit,
II) high mass planet on a low eccentricity orbit,
III) low mass planet, moderate eccentricity orbit, and 
IV) high mass planet, moderate eccentricity orbit.
\label{fig:patternfig}}
\end{figure}

Figure~\ref{fig:patternfig} shows these structures here for reference; 
we discuss them throughout the paper, particularly in Section 4.
Cases I and II in this figure represent the structures formed by planets on 
low-eccentricity orbits.  Cases III and IV represent structures created
by planets on moderately eccentric orbits.
Cases I and III represent structures created by planets with substantially
less than 0.1\% of the mass of the star; cases II and IV represent
structures created by more massive planets.

\section{PLANETS ON LOW-ECCENTRICITY ORBITS}

\subsection{Low Mass Planets}

First, we review some of the physics of resonant dust rings created by
relatively low mass planets, like the Earth and Neptune.
In the rest of this paper, we generalize
the discussion to planets with mass up to $\sim15$ Jupiter masses
and $e_0$ up to $\sim 0.6$.
The resonant geometries we will describe apply to the general restricted
three-body problem.  However, we have in mind a 
a source of dust, like the asteroid belt
or the Kuiper belt, which releases dust on low-eccentricity orbits at
semimajor axes, $a > a_{0}$, where $a$ is the semimajor axis of the
particle's orbit and $a_0$ is the semimajor axis of the planet's orbit.
We use the convention that quantities with a subscript 0 refer to the planet.

We do not consider interactions between the dust grains, or between dust
and any gas in the disk.  For example, our approach does not apply to the
gaseous, optically-thick disks found around young stellar objects.
However, the resonances we describe must underlie the basic resonant features
of the solar system dust complex and debris disks around main
sequence stars with less than a few lunar masses of dust.
  
The radiation forces on a particle are parametrized by $\beta$,
the strength of the stellar radiation pressure force on a particle
divided by the strength of the stellar gravitational force on a particle.
For a spherical particle with radius $s \gtrsim 1$~$\mu$m and density 2 g~cm${}^{-3}$
orbiting a star with mass $M_{\star}$ and luminosity $L_{\star}$, 
\begin{equation}
\beta= \left({{0.285 \ \mu\hbox{m} }\over{s}} \right) \left( {L_{\star} \over L_{\odot} }\right)\left({M_{\odot} \over M_{\star} } \right).
\end{equation}
A typical particle in the solar system may have
$s\approx$1--100~$\mu$m \citep{grun85,fixs02}, or $\beta\approx$0.285--0.00285.
Dust grains released at circumstellar distance $r$ that are too large to be ejected
by radiation pressure ($\beta \lesssim 0.5$) spiral into the star via
P-R drag \citep{robe37, burn79}, on a time scale 
\begin{equation}
T_{PR}={{400} \over {\beta}} \left({M_{\odot} \over M_{\star} } \right) \left({{r}\over{1 AU}} \right)^2 {\rm years}.
\end{equation}
Other drag forces, like solar wind, may also contribute to the decay of
the particle's orbit \citep{bana94}.

Dust spiraling inwards towards a planet encounters
a series of exterior MMRs, each of which is associated with terms
in the disturbing function of the form 
\begin{equation}
<R>_{res}= {{G m_0} \over {a}} F(\alpha, e, e_0) \cos{\phi},
\label{eq:resterm}
\end{equation}
where $G$ is the gravitational constant,
$m_0$ is the mass of the planet, and $\alpha=a_0/a$
(see, e.g., Brouwer \& Clemence 1961; Murray \& Dermott 1999).  The
resonant argument, $\phi$, is a
linear combination of the orbital elements of the particle and the planet,
which can be interpreted as an angle \citep{gree78}. 
We do not discuss interior MMRs because they can not sustain long-term
trapping (see e.g. Murray \& Dermott 1999, p. 381).
The potential, $<R>_{res}$, causes the argument $\phi$ to accelerate.  For
a particle trapped in the resonance, $\phi$ librates about $\phi \approx 0$
if $F(\alpha, e, e_0) < 0$, and $\phi$ librates about $\phi \approx \pi$
if $F(\alpha, e, e_0) > 0$. 
A MMR of the form $j$:$k$ is nominally located at a semimajor axis
given by $1/\alpha \approx (j/k)^{2/3}(1 - \beta)^{1/3}$.

As a particle approaches the planet from afar, it encounters
stronger and stronger resonances, and has a better and
better chance of becoming trapped.
For particles approaching the Earth or Neptune, the first resonances
which are strong enough to trap substantial amounts of dust are the
first-order MMRs: resonances of the form 2:1, 3:2, 4:3, etc. To first
order in eccentricity and inclination,
these resonances consist of pairs of terms with arguments
\begin{mathletters}
\begin{eqnarray}
\phi_1 &=& j\lambda_{} - (j-1)\lambda_{0} - \varpi 
\label{eq:firstordertermsa} \\
\phi_2 &=& j\lambda_{} - (j-1)\lambda_{0} - \varpi_{0},
\label{eq:firstordertermsb}
\end{eqnarray}
\end{mathletters}
where $\lambda$ and $\lambda_0$ are the mean longitudes of the particle and the
planet, and $\varpi$ and $\varpi_0$ are the longitudes of pericenter of the
particle and the planet, respectively.
For the Earth ($e_0=0.017$) and Neptune ($e_0=0.0086$),
the $\phi_1$ resonance dominates at all values of $\varpi$ if the dust particle
has even a small orbital eccentricity, and $\phi_1$ librates about $\phi_1 \approx \pi$. 
Passage through this resonance slowly
raises the particle's eccentricity, which asymptotically
approaches a limiting value, $e_{max}$, when the dynamics are followed using an
expansion to second order in $e$ \citep{weid93, sica93, beau94, liou97}:
\begin{equation}
e_{max}=\sqrt{2/(5j)}.
\label{eq:emax}
\end{equation}
In the process, the
particle's orbit becomes planet crossing, and in a matter of a few
P-R times, the particle generally leaves the resonance
via a close encounter with the planet \citep{marz94} and an abrupt transition
to a new orbit with a different eccentricity and semimajor axis.
Near the Earth, trapped dust populates several first-order MMRs.

\begin{figure}
\epsscale{0.85}
\plotone{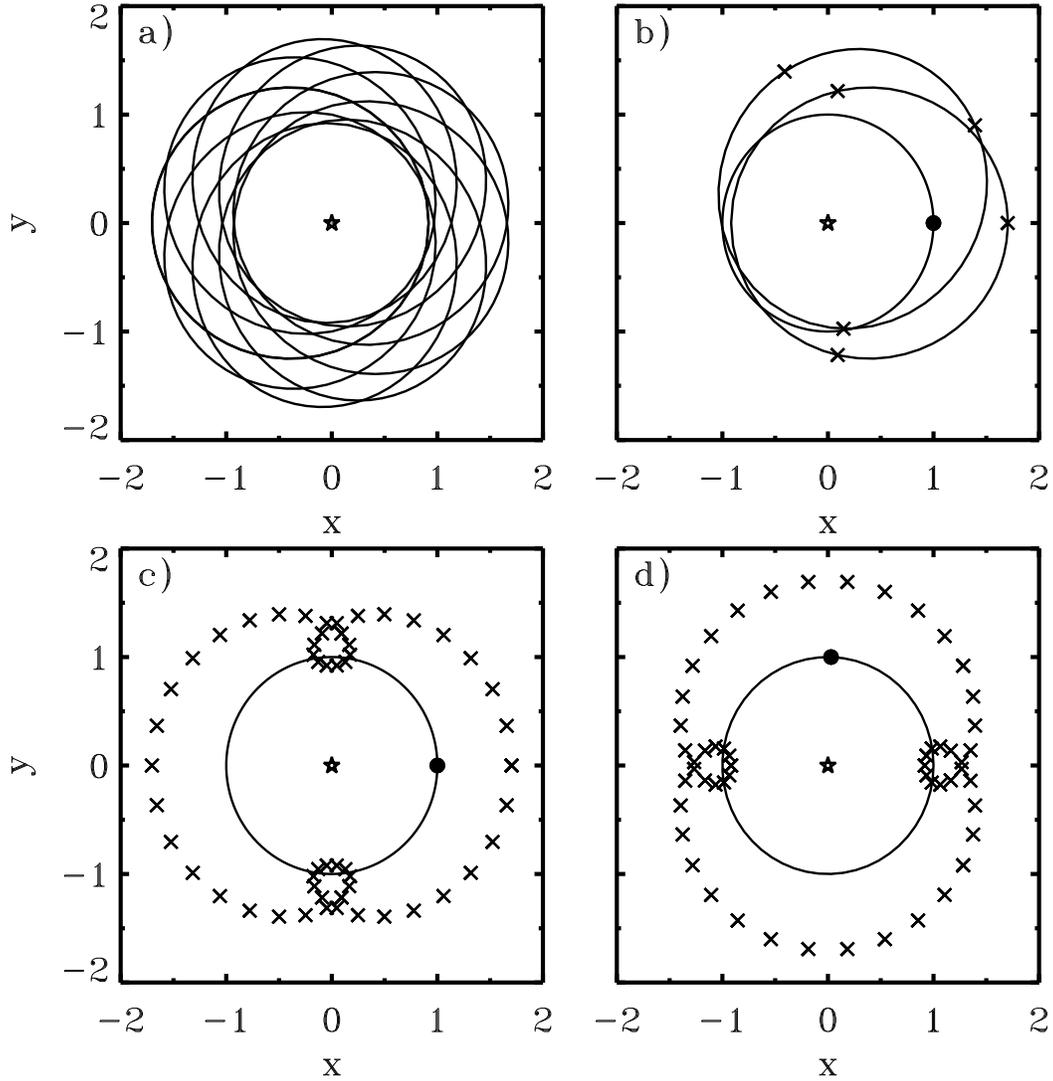}
\figcaption{How a planet on a low eccentricity
orbit creates density waves. a) Several particle orbits with
different $\varpi$'s.
b) Libration centers of the $3 \lambda - 2 \lambda_0 - \varpi$ term
on two of these orbits, indicated by X's.
c) The locus of all the libration centers. 
d) The density wave appears to orbit at the same angular frequency as the planet
(located at the filled circle).
\label{fig:loweres}}
\end{figure}

The condition that  $\phi_1$ librates about $\phi_1 \approx \pi$
provides a relationship between $\lambda$ and $\varpi$.
Since rotations of the whole system by $2 \pi$ must not affect the dynamics, we
can write that the libration centers are located where
\begin{equation}
\lambda \approx ((j-1)(\lambda_{0} + 2 \pi K) + \varpi + \pi)/j  \qquad \hbox{for $K \in {\cal Z}$}.
\label{eq:longitudes}
\end{equation}
Figure~\ref{fig:loweres} shows how this condition leads to the formation
of a density wave, using the 3:2 MMR as an example.
Figure~\ref{fig:loweres}a shows a variety of elliptical
dust orbits, all with the same $1/\alpha=1.31$ and $e=0.8 e_{max}$, but
with different $\varpi$'s.  According to Equation~\ref{eq:longitudes}, each
one of these orbits has $j=3$ different longitudes 
that may be libration centers for a particle trapped in the $\phi_1$ term.
Figure~\ref{fig:loweres}b shows these
three locations, marked with X's, for each of two elliptical orbits with
slightly different $\varpi$'s.

Figure~\ref{fig:loweres}c shows the locus of 
the libration centers for the whole range of elliptical orbits shown in
Figure~\ref{fig:loweres}a.   In Figure~\ref{fig:loweres}d, the planet has
moved to a new longitude, and following Equation~\ref{eq:longitudes}, 
so have the libration centers.  The density wave formed by
the trapped particles resembles the pattern formed by the libration centers,
blurred somewhat by libration, mostly in the azimuthal direction.  \citet{ozer00}
shows some examples of how libration blurs the patterns formed by particles
trapped in individual resonances.

Though any individual particle in Figure~\ref{fig:loweres} moves steadily
counterclockwise, slower than the planet, the density wave
appears to rotate as a fixed pattern
together with the planet. 
At conjunction ($\lambda_{}=\lambda_{0}$) the resonance condition
implies that $\varpi \approx \lambda_0 + \pi$, so the libration centers
just outside the planet are always near apocenter.
This condition creates the signature gap at the location of the planet
seen in simulations of the Earth's ring \citep{derm94} and of Kuiper Belt
dust interacting with Neptune \citep{liou99} and illustrated in
Figure~\ref{fig:patternfig} (Case I).

\subsection{Higher Mass Planets}

We can understand the range of possible resonant dust cloud structures by understanding
the density-wave patterns created by the series of MMRs that are likely to trap
dust near a planet.  Table~\ref{tab:principal}
lists the arguments in a series of MMRs that figure most prominently in
the sculpting of dust clouds (ignoring the planet's inclination, which
first appears with order inclination squared) in order by $\alpha$.
It also lists the corresponding leading terms in $F(\alpha,e,e_0)$,
evaluated for $\mu=1/1047$ (i.e. Jupiter), and $\beta << 1$.  The first MMRs
in the list are the first-order resonances described above.  

\begin{table}[h]
     \caption{Resonant Arguments.}
      \medskip
\begin{tabular}{ccclc} \hline\hline
Resonance & Nominal $1/\alpha$ & $e_{forced}/e_0$ &  Argument                                & Leading Term in $F(\alpha, e, e_0)$ \\
\hline 
6:5 & 1.13 & 0.96 & $6 \lambda - 5\lambda_0 - \varpi_0 $                  & $-4.44181\ e_0$ \\
    &      &      & $6 \lambda - 5\lambda_0 - \varpi   $ & $4.87053 \ e$ \smallskip \\ 
5:4 & 1.16 & 0.94 & $5 \lambda - 4\lambda_0 - \varpi_0 $                  & $-3.64001\ e_0$ \\
    &      &      & $5 \lambda - 4\lambda_0 - \varpi   $ & $4.07424\ e$ \smallskip \\
4:3 & 1.21 & 0.92 & $4 \lambda - 3\lambda_0 - \varpi_0 $                  & $-2.83462 \ e_0$ \\
    &      &      & $4 \lambda - 3\lambda_0 - \varpi   $ & $3.27756 \ e$ \smallskip \\
3:2 & 1.31 & 0.87 & $3 \lambda - 2\lambda_0 - \varpi_0 $                  & $-2.02226 \ e_0$ \\
    &      &      & $3 \lambda - 2\lambda_0 - \varpi   $ & $2.48115 \ e$ \smallskip \\
2:1 & 1.59 & 0.74 & $2 \lambda - \lambda_0 - \varpi_0  $ & $-1.18945\ e_0$ \\
    &      &      & $2 \lambda - \lambda_0 - \varpi    $\tablenotemark{a} & $0.426628\ e$ \smallskip \\
3:1 & 2.08 & 0.58 & $3 \lambda - \lambda_0 -2 \varpi_0 $ & $0.598757 \ e_0^2 $   \\
    &      &      & $3 \lambda - \lambda_0 -\varpi_0-\varpi  $ & $-2.21298 \ e_0e  $  \\
    &      &      & $3 \lambda - \lambda_0 -2 \varpi   $\tablenotemark{a} & $-0.514804\ \ e^2$ \smallskip  \\
4:1 & 2.52 & 0.49 & $4 \lambda - \lambda_0 - 3\varpi_0 $ & $-0.244422 \ e_0^3  $   \\
    &      &      & $4 \lambda - \lambda_0 - 2\varpi_0 - \varpi$ & $1.61636   \ e_0^2e $   \\
    &      &      & $4 \lambda - \lambda_0 - \varpi_0 -2\varpi $ & $-3.51697  \ e_0e^2 $   \\
    &      &      & $4 \lambda - \lambda_0 - 3\varpi           $\tablenotemark{a} & $1.32796\   \ e^3$  \smallskip  \\
5:1 & 2.92 & 0.42 & $5 \lambda - \lambda_0 - 4\varpi_0         $ & $0.0848968 \ e_0^4  $  \\
    &      &      & $5 \lambda - \lambda_0 - 3\varpi_0 - \varpi$ & $-0.855830 \ e_0^3e $   \\
    &      &      & $5 \lambda - \lambda_0 - 2\varpi_0 -2\varpi$ & $3.20820   \ e_0^2e^2 $   \\
    &      &      & $5 \lambda - \lambda_0 - \varpi_0 -3\varpi $ & $-5.28443  \ e_0e^3  $  \\
    &      &      & $5 \lambda - \lambda_0 - 4\varpi           $\tablenotemark{a} & $2.24457\ \ e^4$ \smallskip   \\
\hline\hline
\end{tabular}
\label{tab:principal}
\tablenotetext{a}{Includes contribution from indirect term.}
\end{table}

For a given planet/dust cloud combination, a few
MMRs generally dominate the observed resonant structure.
In simulations of $\beta=0.037$ dust particles approaching the Earth
\citep{derm94}, the first-order $j=$4--10 MMRs dominate the appearance of the
trapped dust cloud.  In simulations of Neptune's ring by \citet{liou99}, the
$2:1$ and $3:2$ resonances ($j=2$ and $j=3$) dominate the appearance
of the dust cloud.

The factor of $Gm_0/a$ in
Equation~\ref{eq:resterm} makes MMRs at large $a$ 
weaker and less able to trap substantial quantities of dust.  But
for more massive planets, the trapping probability for all MMRs
is higher \citep{lazz94}, so particles approaching from afar become trapped
sooner, in MMRs with longer orbital periods and larger semimajor axes.
Likewise, close encounters with the planet more easily
scatter dust grains out of exterior MMRs at small $a$, and this effect
worsens with the mass of the planet.

The MMRs near the planet also become close together and
begin to compete with one another.  If $\mu=m_0/M_{\star}$, the
resonance overlap criterion \citep{wisd80, dunc89}
predicts that first-order resonances with $j > 0.45 (\mu(1-\beta))^{-2/7}+1$ are
completely chaotic.  This condition appears to place an absolute limit
on how far down the chain of resonances a dust particle can be trapped.
For large particles near the Earth ($\mu \approx 3 \times 10^{-6}$), this criterion predicts that
the first completely overlapped MMR is 17:18, for Neptune ($\mu \approx 5 \times 10^{-5}$), 
it is 8:9, and for Jupiter ($\mu \approx 10^{-3}$) it is 4:5.

As we have mentioned, many of the observed extrasolar planets have substantially
more mass than planets in the solar system.  
Such massive planets quickly scatter dust from their first-order MMRs.
So after the first-order resonances,
Table~\ref{tab:principal} lists the $n$:1 resonances, the lowest order
terms available at large semimajor axes---beyond the 2:1 MMR (nominally located
at $a/a_0=1.59$).  In numerical
integrations, \citet{kuch01} and \citet{wiln02} found that dust spiraling
inwards towards a massive planet became trapped in this series of $n$:1
resonances.  Table~\ref{tab:principal} shows that the terms
in these resonances which don't depend on the eccentricity of the planet
have the form $j \lambda - \lambda_0 -(j-1)\varpi$.  These terms must
dominate when $e_0$ is small. 

The $j \lambda - \lambda_0 -(j-1)\varpi$ terms in Table~\ref{tab:principal}
are mostly positive, so most of these arguments librate around $\pi$.  However
these terms all include contributions from indirect terms, which arise from
the reflex motion of the star.  The indirect contributions tend to reduce
the strength of the resonance, and in the case of the 3:1, this effect
makes the coefficient negative, so the $3 \lambda - \lambda_0 - 2 \varpi$
argument librates around $0$. 

The libration centers for the terms which librate around $\pi$ are
located at
\begin{equation}
\lambda   \approx ((j-1)\varpi + \lambda_0 + \pi (1 + 2K))/j, \qquad \hbox{for $K \in {\cal Z}$}.
\label{eq:longitudes2}
\end{equation}
Again there are $j$ libration centers spaced evenly around each elliptical
dust orbit, and again, the locus of libration centers appears to co-rotate
with the planet.  But this time, when we set
$\lambda=\lambda_0$, we find
that at conjunction, the particle can be in one of a few different places;
\begin{equation}
\varpi \approx \lambda_0 - \pi (1+2K)/(j-1)  \qquad \hbox{at conjunction}.   
\end{equation}

One can also show following \citet{weid93}, for example, that passage through a pure
$j \lambda - \lambda_0 -(j-1)\varpi$ term raises the eccentricity of a
dust particle in the same way that passage through a
$j \lambda - (j-1)\lambda_0 -\varpi$ term does, and that the limiting
eccentricity is again given by Equation~\ref{eq:emax}.
We can generate a rough picture of the density waves created by the dust trapped
in these terms in the same manner as Figure~\ref{fig:loweres}, by
drawing a variety of elliptical orbits with eccentricity $e \approx e_{max}$
at an appropriate semimajor axis, and using Equation~\ref{eq:longitudes2}
(and Kepler's equation) to locate the libration centers on these orbits.

\begin{figure}
\epsscale{0.95}
\plotone{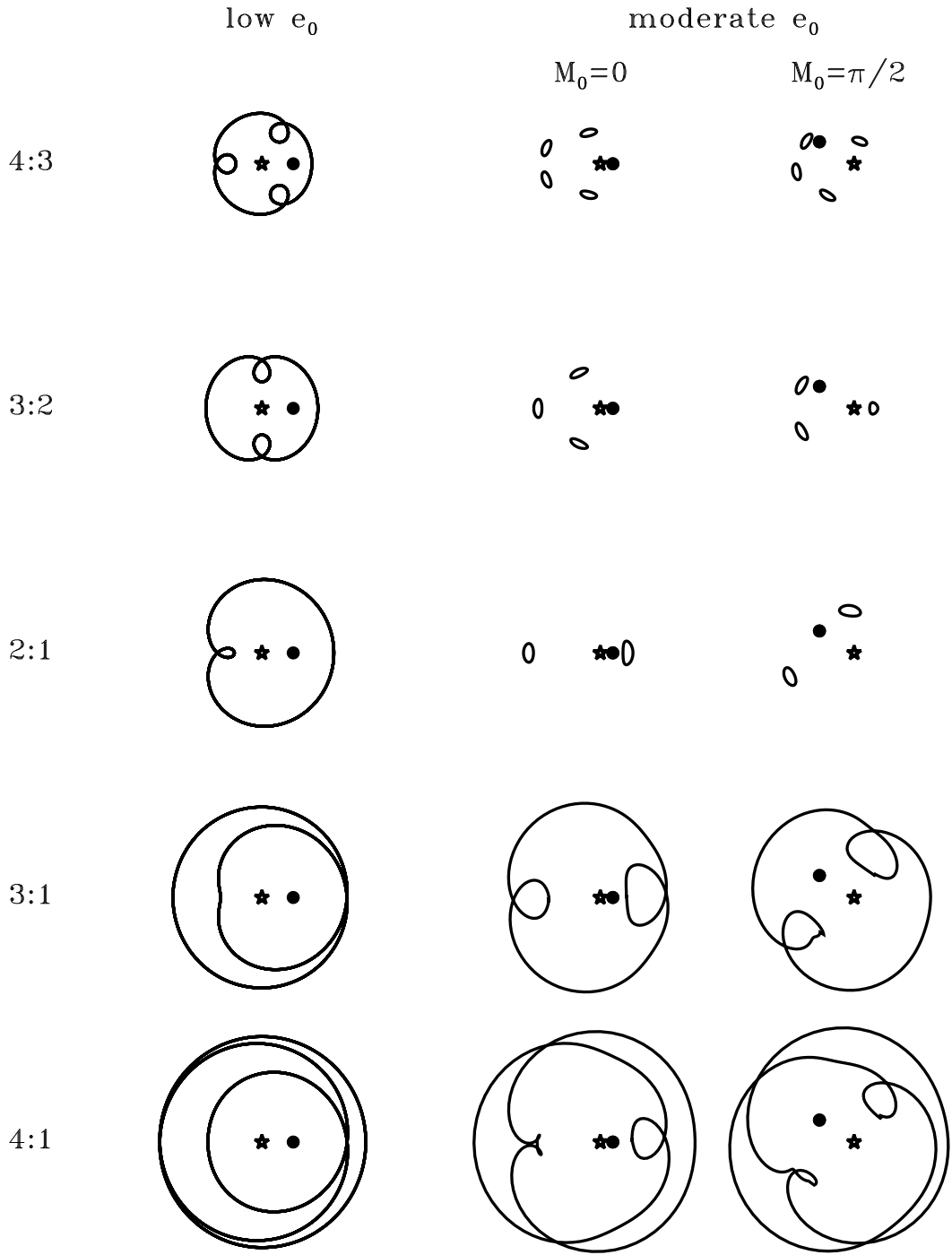}
\figcaption{Patterns formed by dust in MMRs listed in Table~\ref{tab:principal}.
\label{fig:efig}}
\end{figure}

Figure~\ref{fig:efig} shows the patterns created by the most important
resonant terms listed in Table~\ref{tab:principal}.  
The first column of Figure~\ref{fig:efig} shows patterns for the terms which
appear when the planet's eccentricity is low: the
$j \lambda - (j-1) \lambda_0 - \varpi$ terms for first-order resonances, and the
$j \lambda - \lambda_0 -(j-1)\varpi$ terms for $n$:1 resonances.  
The patterns are the locii of the libration
centers, generated in the manner of Figure~\ref{fig:loweres}.

\section{PLANETS ON ECCENTRIC ORBITS}

The patterns in the first column of Figure~\ref{fig:efig} appear in the
textbook by Murray and Dermott (1999, p. 325), derived in a slightly different
context---by tracing the path of a particle in a frame corotating with the
planet.  However, the literature offers little discussion of
resonant structures created by planets on eccentric orbits, and this case is
crucial for understanding extrasolar planetary systems.  Naturally, for a planet
on an eccentric orbit, there is no simple corotating frame.  But we can
deduce the resonance patterns associated with planets on eccentric
orbits by building on the arguments used above.

\subsection{Secular Effects}

Besides resonant perturbations, planets introduce
secular perturbations to the orbits of nearby particles.
When the planet's orbit is eccentric, these perturbations
generally produce a correlation between $e$ and $\varpi$.
We must consider these effects when we examine MMRs with planets on
eccentric orbits.


We can demonstrate the importance of secular perturbations by comparing them
with radiation forces.  Near a planet,
the P-R time, and thereby the trapping time, is longer than the secular
time scale for dust grains with $\beta < \beta_0$, where 
\begin{equation}
\beta_0 =100 \left( {M_{\star} \over M_{\odot}} \right)^{-1/2}
\left( {a \over 1 \hbox{AU}} \right)^{1/2} \alpha \ b^1_{3/2}(\alpha) \mu,
\end{equation}
and the Laplace coefficient (size $\sim$ unity) is 
\begin{equation}
b^m_{3/2}(\alpha)={1 \over \pi}\int_0^{2 \pi} {{\cos{m \psi} \ d \psi} \over{(1-2 \alpha \cos{\psi} + \alpha^2)^{3/2}}} .
\end{equation}
I.e., $\beta_0 \sim 100 \mu$.
For example, $\beta_0=0.022$ for Neptune, $\beta_0=0.44$ for a Jupiter-mass
planet orbiting Vega at $a=40$ AU.  So we can expect most observable particles in the
cases we are interested in ($10^{-5} \lesssim \mu \lesssim 10^{-2}$) to
suffer significant secular evolution while they are trapped in a MMR.
Secular perturbations from planets with eccentric orbits
affect all dust particles in their vicinity, even those which are not
in strong MMRs.

In the Laplace-Lagrange description of secular perturbations, valid for $e,e_0 << 1$,
a particle's osculating $e$ and $\varpi$ are expressed
as a combination of a constant forced elements, $e_{forced}$ and
$\varpi_{forced}$, and time-varying free elements, $e_{free}$
and $\varpi_{free}$.
Secular evolution is easily visualized in the $(h,k)$ coordinate system,
where $h= e \cos{\varpi}$ and $k=e \sin{\varpi}$.
In this system, we write $h=h_{forced}+h_{free}$, and $k=k_{forced}+k_{free}$.

When there is only a single perturbing planet on a fixed orbit, the particle's
forced elements are constant, and the osculating elements, $h$ and $k$,
trace out a circle centered on $(h_{forced},k_{forced})$
with a radius of $e_{free}$.  The forced longitude of pericenter is
$\varpi_{forced}=\arctan(k_{forced}/h_{forced})=\varpi_0$, and the
forced eccentricity is 
\begin{equation}
e_{forced}
= \sqrt{h_{forced}^2 + k_{forced}^2}
={{b_{3/2}^{2}(\alpha)} \over {b_{3/2}^{1}(\alpha)}} e_0 .
\end{equation}
As $a$ approaches $a_0$, $e_{forced}$ approaches $e_0$.
Table~\ref{tab:principal} lists the approximate value of $e_{forced}/e_0$ at the
nominal $\alpha$ for each MMR.

In a cloud of many particles,
many orbits with a range of $h_{free}$ and $k_{free}$ are occupied.
However, all particles with a given semimajor axis will have the same $h_{forced}$
and $k_{forced}$.  For example, if all particles are released on circular
orbits outside the planet's orbit and outside of any MMRs, then the Laplace-Lagrange
solution prescribes that at any given time all the
particles with a given semimajor axis will occupy a circle with
radius $e_{free}$ centered on the point $(h_{forced},k_{forced})$ in
the $(h,k)$ plane.  \citet{derm85} and \citet{wyat99} showed that a set
of orbits occupying such a circle in the ($h,k$) plane form a cloud
that is roughly circular, but
the center of the cloud is offset from the star a distance $e_0 a_0$
along the planet's apsidal line in the direction of the center of the planet's orbit.
Hence, the background dust cloud in the vicinity of a large planet
should often appear circular in the absence of MMRs, though if the
planet's orbit is eccentric, the center of the circle
will be offset from the star.

When the perturber has higher eccentricity, or when a particle is 
in a MMR, the particle's secular trajectory changes
somewhat; the orbit no longer traces an exact circle in the $(h,k)$ plane,
even when the librations are averaged away \citep{wisd83}.
However, the character of the averaged secular motion often remains the same;
$h$ and $k$ follow a simple closed loop around $(h_{forced},k_{forced})$.  
We will retain the spirit of the Laplace-Lagrange approximation
for the secular motion of the particles for the remainder of this paper,
and we wil refer loosely to free and forced elements, even when we are
discussing resonant orbits.  We find that this approximation suffices for
our broad exploration of planetary signatures in debris disks.

\subsection{Mean Motion Resonances}

When a particle is trapped in a MMR with a planet on a circular orbit,
the resonant argument which only depends on $\varpi$, not on 
$\varpi_0$ dominates the particle's motion.  The other terms in
the resonance have coefficients $F(e,e_0)$ that are zero when $e_0$ is zero.
However, when a particle is trapped in a MMR with a planet on
an eccentric orbit, resonant arguments which depend on $\varpi_0$
may come into play. 

To leading order, the averaged disturbing function for a MMR with a planet on an
eccentric orbit is the sum of two or more terms:
\begin{equation}
<R>_{res}={{G m_0} \over a }\sum_{\xi} F_{\xi}(\alpha, e, e_0) \cos{\phi_{\xi}},
\end{equation}
where, as Table~\ref{tab:principal} shows, $\phi_1=p \lambda - q \lambda_0 + (p-q) \varpi$
and $\phi_{\xi}=\phi_1+({\xi}-1)(\varpi-\varpi_0)$.
Following \citet{wisd83}, we can re-express this sum as
\begin{equation} 
<R>_{res}= {{G m_0} \over a } F'(\alpha, e, e_0, \varpi, \varpi_0) \cos{\phi'},
\end{equation}
where 
\begin{equation}
\phi'=\phi_1+\arctan\left({{\sum_{{\xi}=2}^{p} |F_{\xi}| \sin\left(({\xi}-1)(\varpi -\varpi_0) + \delta_{\xi}\right)}
\over {|F_{\xi}| + \sum_{{\xi}=2}^{p} |F_{\xi}| \cos\left(({\xi}-1)(\varpi -\varpi_0)  + \delta_{\xi}\right) }}\right)
\label{eq:phiprime}
\end{equation}
and 
\begin{equation}
F'(\alpha, e, e_0, \varpi, \varpi_0)=
\sqrt{
\left(\sum_{\xi} F_{\xi}(\alpha, e, e_0) \sin{\phi_{\xi}} \right)^2
+ \left(\sum_{\xi} F_{\xi}(\alpha, e, e_0) \cos{\phi_{\xi}} \right)^2
}.
\label{eq:fprime}
\end{equation}
The quantity $\delta_{\xi}=\pi$ for $F_{\xi} <0$, and $\delta_{\xi}=0$ otherwise.
We define $F'$ to be always $\ge 0$, so on resonance, the new argument, $\phi'$,
librates around $\phi' \approx \pi$. 
Equation~\ref{eq:phiprime} shows explicitly that the differences among
the terms only appear on a secular time scale, as $\varpi-\varpi_0$ precesses.

At any moment, the particle's orbit may be viewed as undergoing libration 
about $\phi'=\pi$.  This change in the effective resonant argument can
change the constraints on the orbital elements of resonant objects, which
can result in dramatically different-looking clouds of trapped particles.
Often one term dominates, in the sense that
$\phi' \approx \phi_{\xi}$ or $\phi' \approx \phi_{\xi}+\pi$.  This circumstance depends on
$e$, $e_0$, $\varpi$, and $\varpi_0$, and the details of trapping,
but we can appeal to numerical simulations and use what we know about
the secular evolution of dust particles to help decide which terms are
likely to be most important.

\subsection{High Mass Planets}

It is easier to consider high mass planets first, because the $n$:1 resonances
that dominate the appearance of dust clouds
containing a massive planet are farther from the planet than the
first-order resonances which dominate in the case of a low mass planet.
Particles released on circular orbits approach a planet with
$e_{free} \approx e_{forced}$, so they reach the $n$:1 MMRs
with $e$ ranging from 0 to $e_{forced}+e_{free}$.
As Table~\ref{tab:principal} shows, $e_{forced} \approx e_0/2$ near
the lowest order $n:1$ MMRs.  When $\varpi \approx \varpi_0+\pi$, all
the terms become degenerate, in the sense that all of their libration
centers occur at the same longitudes.  At the other end of the range
of secular motion, $\varpi \approx \varpi_0$, and $e \approx e_0$, so
according to Equation~\ref{eq:phiprime}, the term with the largest
coefficient in $F(\alpha, e, e_0)$ will dominate the motion of the particles.

The arguments which match the terms with the
largest coefficients in $F(\alpha, e, e_0)$ (see Table~\ref{tab:principal})
share a second common form: $j \lambda - \lambda_0 -\varpi_0 -(j-2)\varpi$.  
Numerical integrations \citep{wiln02} confirm
that these terms to dominate
the resonant dynamics over a wide range of planet eccentricities.
These terms always have $F(\alpha, e, e_0) < 0$, so the corresponding
arguments librate about 0.  The $j$ libration centers on each ellipse
are located at
\begin{equation}
\lambda \approx (\lambda_0 +\varpi_0 +(j-2)\varpi + 2 \pi K)/j, \qquad \hbox{for $K \in {\cal Z}$}.
\label{eq:lambda3}
\end{equation}
We reserve a discussion of the complicated secular dynamics of these
resonances for a future paper.

\begin{figure}
\epsscale{0.87}
\plotone{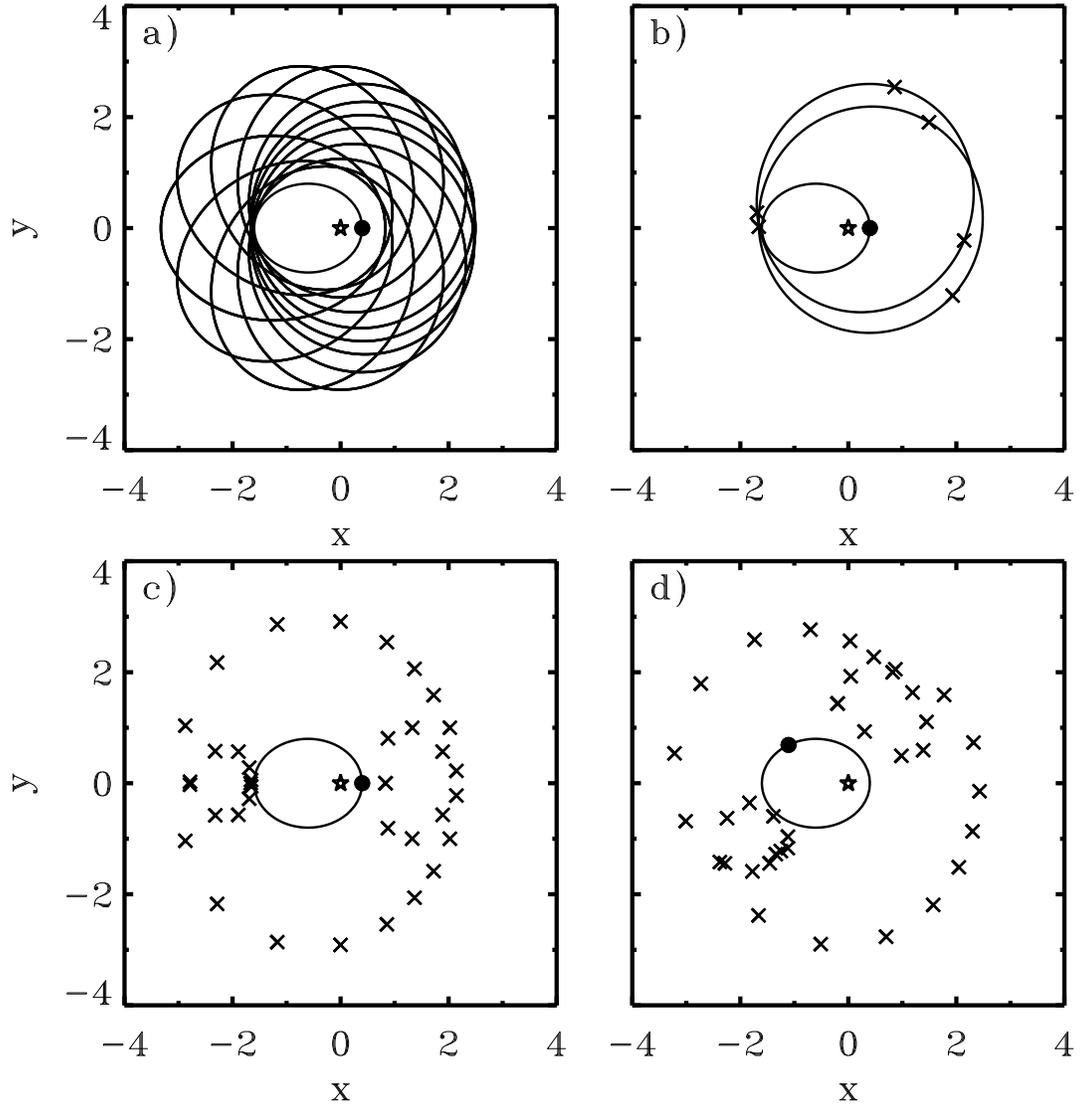}
\figcaption{Libration centers of the
$3 \lambda - \lambda_0 - \varpi_0 - \varpi$ term.
a) Several particle orbits with different $e$ and $\varpi$.
b) The libration centers on two of these orbits when the planet is at pericenter.
c) All the libration centers.
d) The clumps formed by particles trapped in this term appear to rotate at
half the angular frequency of the planet.
\label{fig:whyblob}}
\end{figure}

Figure~\ref{fig:whyblob} illustrates the pattern formed by particles in
the $3 \lambda - \lambda_0 -\varpi_0 - \varpi$ term near a planet with
$e_0=0.6$.  Once a particle is trapped in resonance, its secular evolution
no longer obeys the Laplace-Lagrange solution, but the libration centers of the
trapped particles will still occupy a closed curve in the $(h,k)$ plane.
Figure~\ref{fig:whyblob}a shows such a variety of orbits
centered on $h=0.58 e_0$, $k=0$.
Orbits which are close to apse-alignment with the planet have higher
eccentricities.  Like the Laplace-Lagrange orbit distributions discussed by
\citet{derm85} and \citet{wyat99}, these orbits taken together
form a pattern which is roughly circular, but the center of the circle
is offset from the star along the planet's
apsidal line in the direction of the center of the planet's orbit.

Figure~\ref{fig:whyblob}b shows the 3 libration centers on two of these orbits,
calculated from Equation~\ref{eq:lambda3}.
By definition, $\lambda=M+\varpi$, where $M$ is the particle's mean anomaly.
For particles at pericenter, $M=0$ and $\lambda=\varpi$, so
for particles trapped in {\it any} $j \lambda - \lambda_0 -\varpi_0 -(j-2)\varpi$
term, 
\begin{equation}
\varpi \approx M_0 /2 +\varpi_0 + \pi K  \qquad \hbox{for $K \in {\cal Z}$},
\end{equation}
where $M_0$ is the planet's mean anomaly.  In other words, the libration centers
for this family of terms reach pericenter at two different longitudes, and these
special longitudes precess at an angular frequency equal to half the
Keplerian angular frequency of the planet.  When the planet reaches pericenter
($M_0=0$), so do the particles on orbits that are apse-aligned or anti-apse-aligned
with the planet's orbit.  Figures~\ref{fig:whyblob}c and d show
that the locus of libration centers makes a characteristic two-lobed pattern,
which appears to rotate at half the mean angular frequency of the planet.  The
center of this apparent rotation lies on the planet's apsidal line between the star and
the center of the planet's orbit.

\subsection{Low Mass Planets}
\label{sec:lowmassplanets}

\begin{figure}
\epsscale{0.85}
\plotone{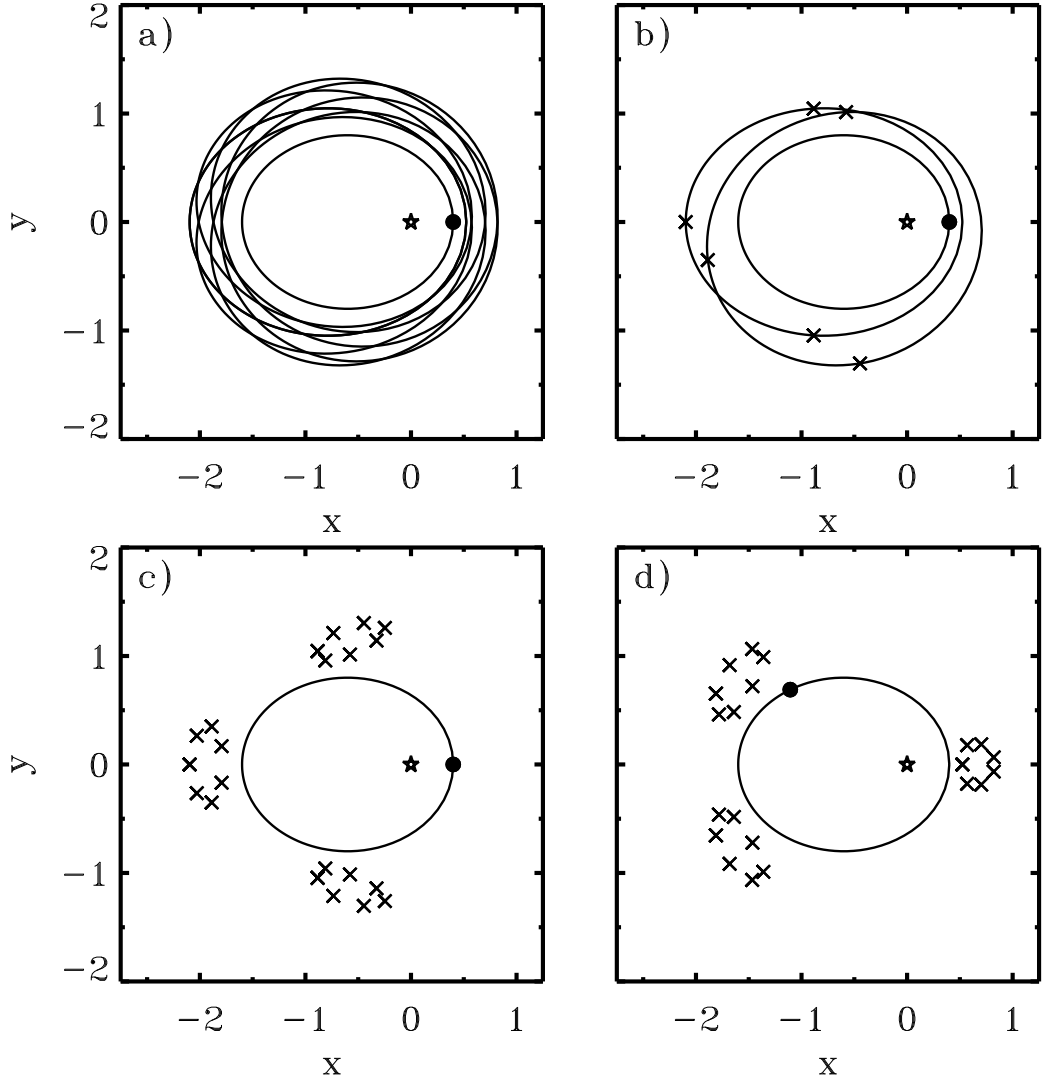}
\figcaption{How dust confined to the 3:2 MMR near a planet on a
moderately eccentric orbit might appear.
a) Several orbits with different $e$ and $\varpi$, distributed in a
Laplace-Lagrange circle with small $e_{free}$.
b) The libration centers on two of these orbits when the planet is at pericenter.
c) All the libration centers.
d) The three dust clumps really do orbit at 2/3 of the planet's orbital frequency.
\label{fig:whybloblowmass}}
\end{figure}

The resonances populated near low-mass planets cluster at semimajor axes
near the planet's semimajor axis, where $e_{forced} \approx e_0$.
In this vicinity, the resonant orbits can easily
become planet-crossing unless they are roughly apse-aligned with the planet's
orbit, i.e. $e_{free} < e_{forced}$, and $\varpi \approx \varpi_0$.
Figure~\ref{fig:whybloblowmass}a shows a collection of roughly apse-aligned
orbits, all with $e_{forced}=0.5$, and $e_{free} = 0.12$. In the absence of MMR's,
the combination of these elliptical orbits would appear as an elliptical ring.

For the case of the 3:2 MMR,
a single one of these orbits has three libration centers,
shown in Figure~\ref{fig:whybloblowmass}b.
For a family of orbits that must remain roughly apse-aligned with the planet,
the secular spread in the libration centers is necessarily
small, and the resonant arguments are all roughly equal. 
So, as Figure~\ref{fig:whybloblowmass}c shows, we should expect the
density wave to resemble the pattern
produced by the libration centers in any one orbit; the libration
centers associated with any MMR of the form
$j$:$k$ form a set of $j$ clumps.  As Figure~\ref{fig:whybloblowmass}d shows,
the locus of all the libration centers orbits at $k/j$ times
the angular frequency of the planet; when the planet's mean anomaly changes by
$\pi/2$, the mean anomalies of the clumps of particles in the 3:2 MMR change
by $\pi/3$.

An exterior first-order MMR has two relevant terms
(Equations~\ref{eq:firstordertermsa} and \ref{eq:firstordertermsb}), and for 
the second term, $F_2(\alpha, e, e_0) < 0$.  For this case,
Equation~\ref{eq:fprime} becomes
\begin{equation}
\phi'=\phi_1+\arctan\left({{|F_2(\alpha, e, e_0)| \sin(\varpi-\varpi_0 + \pi)}
\over {|F_1(\alpha, e, e_0)|+[F_2(\alpha, e, e_0)|\cos(\varpi-\varpi_0 + \pi) }}\right).
\label{eq:phiprimefirstorder}
\end{equation}
A key difference between the two terms 
is that one argument librates around 0, the other around $\pi$.

Deciding which resonances dominate in the vicinity of a low mass planet on
an eccentric orbit can be tricky.
When $|F_1(\alpha, e, e_0)| >> |F_2(\alpha, e, e_0)|$, $\phi' \approx \phi_1$.
When $|F_1(\alpha, e, e_0)| << |F_2(\alpha, e, e_0)|$, $\phi' \approx \phi_2 + \pi$.
However, when $F_1(\alpha, e, e_0) \approx F_2(\alpha, e, e_0)$, the two terms tend to
cancel each other, that is, $|F'| \approx |F_1 -F_2|$ when $\varpi \approx \varpi_0$.
This cancellation diminishes the strength of the resonance.

So we should expect first-order resonances to be relatively weak for a dust particle
whose orbit has been secularly apse-aligned with the planet's orbit. 
At $e=e_{forced}$, the $\phi_1$ resonance generally has the larger coefficient.
The 2:1 MMR is an exception, because $F_1$ for the 2:1 is diminished by an
indirect term, so the $\phi_2$ resonance easily dominates.  
We leave it to numerical studies \citep{quil02} to decide which MMRs and terms dominate
in a given system containing a low mass-high eccentricity planet.
However, no matter which MMRs dominate, we expect the same generic behavior;
the trapped dust will form an eccentric ring of dust clumps.

\section{PLANETARY SIGNATURES}

\subsection{Four Basic Cloud Structures}
\label{sec:fourbasic}

The first column of Figure~\ref{fig:efig} illustrates the density waves formed
by the resonant terms we expect will dominate the appearance of dust clouds
near planets with low orbital eccentricity.
The second and third columns illustrate the density waves formed
by the resonant terms we expect will dominate near
planets with moderate orbital eccentricity:
the $j \lambda - (j-1) \lambda_0 - \varpi_0$
terms for first-order resonances, and the
$j \lambda - \lambda_0 -\varpi_0 -(j-2)\varpi$
terms for $n$:1 resonances.  In the second column, the planet is at pericenter;
in the third column, the planet is at $M_0=\pi/2$.

Our map of these basic patterns prepares us to
consider more generally the appearance of a dust disk in the vicinity
of a planet, where trapped dust occupies several MMRs.
Now we can describe the origin of the structures shown in
Figure~\ref{fig:patternfig}; they are superpositions of
the patterns shown in Figure~\ref{fig:efig}.

Cases I and II are superpositions of the patterns depicted in the 
left-hand column of Figure~\ref{fig:efig}.
Cases III and IV are superpositions of the patterns depicted in the
right-hand column of Figure~\ref{fig:efig} (and also in the middle column, at a
different planetary orbital phase).  In general, the rings created by the
low-eccentricity planets appear to co-rotate with the planet, but
the resonant structures created by moderate-eccentricity planets do not,
because the density wave patterns associated with the
resonances they excite vary with the planet's orbital phase.

\begin{trivlist}
\item{\bf Case I:} A low mass planet
with low orbital eccentricity, like the Earth or Neptune, traps dust in
first-order $j \lambda - (j-1) \lambda_0 - \varpi$
resonances.  Case I in shows a superposition of patterns
produced by terms of this form, the patterns in the upper
left-hand column of Figure~\ref{fig:efig}.
Populating these resonances creates a ring
with a gap at the location of the planet.

\item{\bf Case II:} A higher mass planet on a low-eccentricity
orbit traps dust in more distant
$n$:1 resonances, in terms of the form
$j \lambda - \lambda_0 - (j-1) \varpi$.
Case II shows superpositions of patterns produced by such terms,
the patterns in the lower left-hand column of Figure~\ref{fig:efig}.
These resonances create a larger ring with a smooth central hole.

\item{\bf Case III:}  A low mass planet on a moderately eccentric orbit
traps dust in MMRs with small
secular motion near apse-alignment with the planet, creating a blobby
eccentric ring.  The blobs in this ring
appear to be continually created and destroyed, as dust clumps occupying
different MMRs pass through one another.  The more highly concentrated clumps
are located near the apocenter of the planet's orbit, where all the
particles in the $j \lambda - (j-1) \lambda_0 - \varpi_0$ terms
(except for the 2:1) are constrained to have their conjunctions.
The planet may or may not be located near a gap in the ring.

\item{\bf Case IV:} A high mass planet on a moderately eccentric orbit
creates a ring offset from the star containing a pair of
clumps, where the two-lobed patterns of all
the $n$:1 resonances of the form
$j \lambda - \lambda_0 - (j-2) \varpi - \varpi_0$
conincide.   The clumps appear to orbit
a point along the planet's apsidal line, between the star and the planet,
at half the mean orbital frequency of the planet.  

\end{trivlist} 

Naturally, intermediate cases will result in more than one
variety of resonance being populated by dust particles; these clouds 
can possess features of more than one of the extreme cases shown in 
Figure~\ref{fig:patternfig}.
For example, an intermediate mass planet on a low eccentricity orbit 
will have a slightly larger ring with a less prominent gap
at the location of the planet than Case I.  A planet
with intermediate eccentricity may have both a ring component and
blobs, though librational motion may smear these blobs so
that they blend into the ring.

\subsection{Strong Drag Forces; Wakes}

When the drag force is large and the planet's mass is small, the libration centers
can shift substantially from $\pi$.  
For a MMR containing a collection of resonant arguments of
the form $\phi=j \lambda - k \lambda_0 - (j-1)\varpi - ({\xi}-1)(\varpi_0 - \varpi)$,
the resonant perturbations are
\begin{eqnarray}
\left[{{da} \over {dt}}\right]_{Res} &=& -2 j a n \mu F'(\alpha, e, e_0, \varpi, \varpi_0) \sin \phi' \\
\left[{{de} \over {dt}}\right]_{Res} &=& -n \mu F'(\alpha, e, e_0, \varpi, \varpi_0) \sin \phi',
\end{eqnarray}
where the particle's angular frequency, $n$, on resonance is nominally
\begin{equation}
n=(GM_{\star} (1-\beta)/a^3)^{1/2}.
\end{equation}
PR drag causes relatively slow changes in $a$ and $e$ \citep{wyat50}:
\begin{eqnarray}
\left[{{da} \over {dt}}\right]_{PR} &=& {{-GM_{\star}} \beta \over {ac}} {{(2+3e^2)}\over{(1-e^2)^{3/2}}}\\
\left[{{de} \over {dt}}\right]_{PR} &=& {{-5GM_{\star}} \beta \over {2a^{2}c}} {{(2+3e^2)}\over{(1-e^2)^{1/2}}}
\end{eqnarray}
where $c$ is the speed of light.
To locate the libration centers, we set
$\left[{{da} / {dt}}\right]_{PR} + \left[{{da} / {dt}}\right]_{PR} = 0$, and find
\begin{equation}
\sin \phi_0 = -{(GM_{\star}/a)^{1/2} \beta (1- \beta)\over {2 j \mu c F'(\alpha, e, e_0, \varpi, \varpi_0)}} {{(2+3e^2)}\over{(1-e^2)^{3/2}}}.
\end{equation}
On resonance, $\phi'$ librates around $\phi_0$, which is greater than $\pi$.

For the case of a planet on a circular orbit, the observable effect on a dust
cloud is a shift in the locations of the pericenters of the orbits
of the trapped particles.  In the absence of this effect,
the libration centers first reach pericenter an
angle $\pi / k$ behind the planet.
With this effect, the first pericenter is located closer to the planet, at
an angle $(2 \pi - \phi_0)/ k$.  This asymmetry concentrates the
trapped dust from the Earth's ring into a blob trailing the Earth, sometimes
called the Earth's ``wake'' \citep{derm94,reac95}.

\begin{figure}
\epsscale{0.9}
\plotone{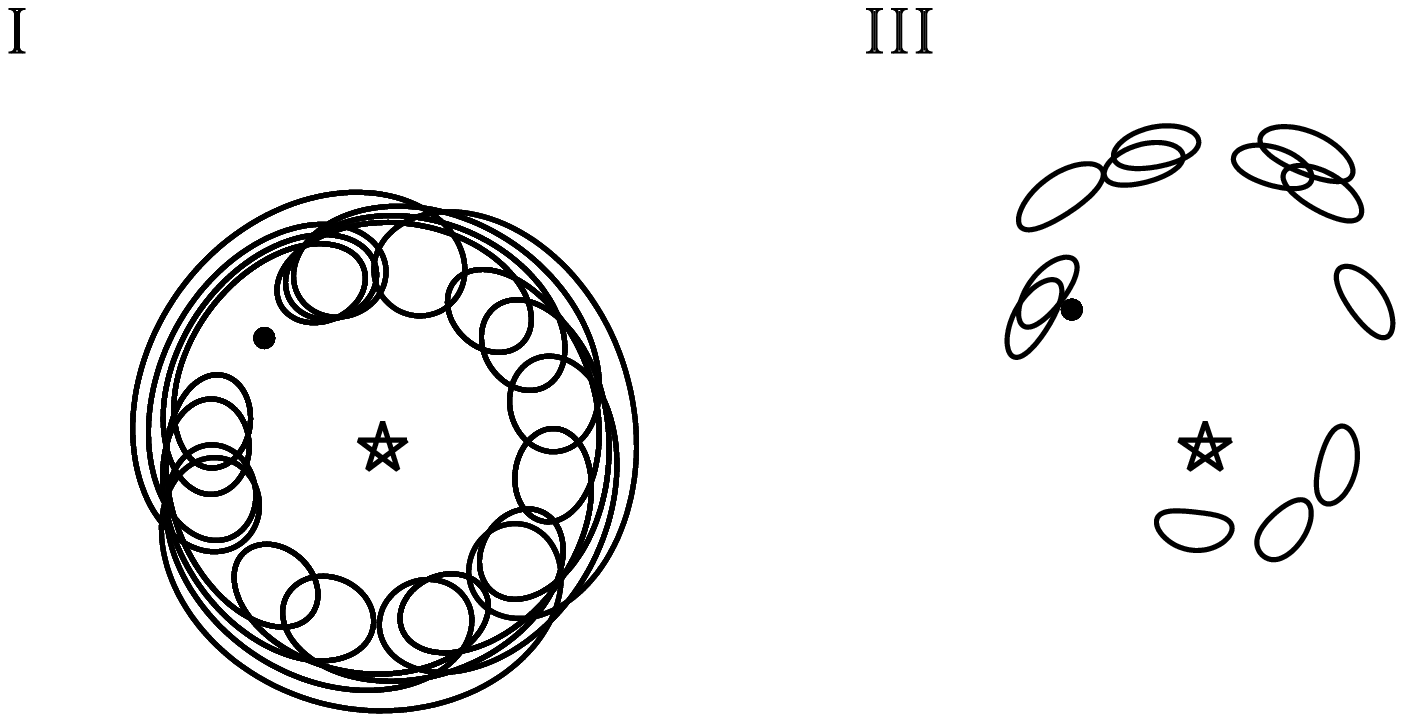}
\figcaption{Cases I and III when the drag force is particularly
strong and the planet's mass is small. Case I develops
a ``wake'' trailing the planet.  The blobs in Case III shift in the prograde direction.
\label{fig:wake}}
\end{figure}

Figure~\ref{fig:wake} shows an example of how cases I and III, as illustrated
in Figure~\ref{fig:patternfig}, could
appear when the perturber's gravity is weak compared to the drag force
and the shifts in the libration centers are substantial.
Case I shows a wake trailing the planet.
For the terms illustrated in columns two and three of Figure~\ref{fig:efig}, the shift in
the libration center appears as a displacement of the density enhancements.
The $n$:1 resonance clumps will appear at
$\varpi = K\pi + M_0/2 + \varpi_0 + (\phi_0 - \pi)/2$, i.e. their longitudes
shift by $(\phi_0 - \pi)/2$.
The first-order resonance clumps near low-mass
planets on moderately eccentric orbits (Case III) 
shift by $(\phi_0-\pi)/k$ in the prograde direction.

This effect depends on the velocity of the dust ($\sim (G M_{\star}/a)^{1/2}$),
on $\beta$, and on the mass of the planet; it is more
pronounced for small dust grains and low-mass planets close to massive and
luminous stars.  It is negligible for patterns generated by more massive
planets (cases II and IV), even if the planets are in 3-day orbits.
Moreover, this effect does not appear in simulations of Neptune's
ring \citep{liou99} since even Neptune has
high mass (and low orbital velocity) compared to the Earth.
The $\epsilon$~Eridani ring seen by the JCMT \citep{holl98}
probably does not show this effect either for the same reasons.

\section{DISCUSSION}

\subsection{Observed Examples}

Earth is the only planet in the solar system whose resonant ring has been
detected \citep{kuch00}.  This wake in this ring
produces an asymmetry in the thermal infrared background between
the direction leading the Earth and the direction
trailing the Earth, which IRAS and COBE DIRBE detected \citep{reac91, derm94, reac95}.
Indeed, the Earth's ring is the only observed resonant dust cloud
structure---in the solar system or elsewhere---identified
with an independently observed planet!  However, models of Kuiper belt
dust interacting with the outer planets suggest that Neptune also creates
a type I ring with a characteristic hole at the location of the planet.

Images of nearby stars with debris disks supply further examples
of rings which point to undetected planets.
Dust clouds around Vega, $\beta$ Pictoris, $\epsilon$ Eridani and
Fomalhaut have all been imaged at submillimeter wavelengths
\citep{holl98, grea98, fomalhaut}.  The $\beta$ Pictoris disk is edge-on,
making its resonant rings hard to distinguish---if it has any.  However,
the other systems appear to contain resonant rings which may be easier to classify.

\begin{trivlist}
\item{\bf Vega:}
The Vega debris disk may provide an example of case IV.
Submillimeter images of Vega \citep{wiln02} show two concentrations of emission that
resemble those in Figure~\ref{fig:patternfig}; they are not
co-linear with the star, and one is closer to the star than the other.
\citet{wiln02} used this model to estimate that the perturbing planet
has mass $\mu \approx 10^{-3}$ and eccentricity $e \approx 0.6$, and used the
locations of the two knots of emission to infer the planet's current loation.

\item{\bf $\epsilon$ Eridani:}
Case III may be a good model for the $\epsilon$ Eridani dust ring,
which has roughly four major concentrations of emission irregularly placed
around the ring, and an apparent gap in the ring.
As Figure~\ref{fig:patternfig} shows, the low-mass, moderate eccentricity
planet model naturally explains the presence of a few dust clumps of
various concentrations.  
\citet{quil02} have developed a numerical model of type III for the
$\epsilon$ Eridani dust ring using the numerical technique developed by
\citet{kuch01}.  They find that dust released just outside
a planet with eccentricity 0.3 and mass $\mu=10^{-4}$ becomes trapped
in the 5:3 and the 3:2 MMRs with the planet, and forms a blobby
eccentric ring like the one in Figure~\ref{fig:patternfig}.
The second order 5:3 MMR may be relatively strong because the
first order resonances suffer from the cancellation described in
Section~\ref{sec:lowmassplanets}.
Though dust trapped in these resonances probably forms relatively
narrow rings, the large beam of the JCMT could easily blur such a
narrow blobby ring into something resembling the \citet{holl98} image,
as \citet{quil02} illustrate.
  
\item{\bf Fomalhaut:}
The A3 V star Fomalhaut has a circumstellar dust cloud,
recently identified as ring with an azimuthal asymmetry representing
$\sim 5$\% of the total flux from the disk \citep{fomalhaut}.  This ring
is difficult to classify, since it is roughly $20^{\circ}$
from edge-on.  \citet{wyat02} suggest that the dust
may trapped in a 2:1 MMR with a planet, and \citet{fomalhaut} suggest
that the dust might be trapped in a  1:1 MMR with a planet.
However, the 1:1 MMRs are much harder to populate with dust than the MMRs we consider
here, and contrary to \citet{ozer00}, we consider it improbable that a single
MMR, even the 2:1, could dominate the appearance of a real dust
cloud, particularly one as collisional as Fomalhaut's.
More likely, the structure in the Fomalhaut ring arises from collisions
\citet{wyat02} or if resonant trapping is responsible for the
azimuthal structure, there is another clump to be found, perhaps
obscured by the limb-brightening.

\end{trivlist}

We know no observed examples of case II rings.  However, 
the close-in extrasolar planets with periods less than $\sim 30$ days
have nearly circular orbits, probably due to tidal effects \citep{marc00}.
Any dust disks associated with these planets could form examples
of case II.

\subsection{Other Considerations}

We have only addressed the geometry of resonant dust orbits.
A variety of other phenomena may
affect the appearance of an actual debris disk.
For example, we have not discussed how the resonances
are populated.  Different resonances may dominate
when a disk is fed by a belt of asteroids or comets whose orbits are 
restricted to a small region of phase space.

Furthermore, many known dust disks are collisional.
For systems like the rings of Saturn, where each particle
has a collision roughly once per orbit, the collisions dominate
the resonant effects we catalog here, and a fluid description of the
particles becomes more appropriate.
Collisions which occur on intermediate time scales, however, 
may leave particles trammelled in the net of the underlying
strong resonances, altering only the spectrum of populated MMRs.
For example, mutual collisions among dust
grains may destroy dust particles before they can access all
the strong resonances; this effect would preserve the basic
four structures shown in Figure~\ref{fig:patternfig}.

We have also neglected the terms in MMRs that depend
on the inclinations, and restricted our analysis to planets
with moderate eccentricity.  At high planet inclinations
($i^2 \sim 1$) and eccentricities ($e^2 \sim 1$), many new
terms in the disturbing function become relevant.
These effects can alter the basic resonant geometries.

Many planetary systems have more than one planet;
we have also only considered the effects of one planet.
Since we are only interested in high-contrast structures,
we may justify this approach by saying we only care about
the first massive planet to encounter the inspiraling dust---massive
enough to create a high-contrast structure by
ejecting most of the dust grains as they pass.
For example, in simulations by \citet{liou99}, Neptune ejects most of
the inspiraling Kuiper belt particles before they can encounter
any other planets; we are interested in analogous planets.  However,
the secular evolution of a multiple-planet system may affect
how a planet interacts with a dust cloud, even if little dust
reaches most of the planets, and they do not carve their own signatures.

Finally, we have tacitly assumed that the dust cloud is 
observed face on.  Disks which are tilted from face-on may show
a variety of asymmetries due to effects other than resonant trapping, such
as the IRAS/DIRBE dust bands \citep{haus84}, secular warps
\citep{auge01}, and limb brightening.

\section{CONCLUSIONS}

Four basic structures probably represent the range of high-contrast
resonant structures a planet with eccentricity 
$\lesssim 0.6$ can create in disk of dust released on
low-eccentricity orbits: a ring with a gap
at the location of the planet, a smooth ring,
a blobby eccentric ring, and an offset ring plus a pair of clumps.
Some of these structures have slowly become revealed in numerical
simulations of particular debris disks; we have chased them to their
dens in the resonant landscape of the 3-body problem.  The crude key we
have assembled should help classify the debris disk structures seen by
upcoming telescopes like SIRTF, SOFIA, ALMA, JWST and Darwin/TPF.

Observing one of these structures instantaneously should allow us to categorize
the planet as high or low mass (compared to Jupiter orbiting the Sun), and low or
moderate eccentricity (compared to $e_0 \sim 0.2$). 
In the case of a ring with a gap or an offset ring plus a pair of clumps,
the image of the face-on cloud directly indicates the current location of the planet
and points to its longitude of perihelion.  In the case of a blobby eccentric ring,
numerical modeling can potentially reveal these parameters.

\acknowledgements

We thank Tommy Grav, Sean Moran and Mike Lecar for helpful discussions.
This work was performed in part under contract with the Jet Propulsion 
Laboratory (JPL) through the Michelson Fellowship program funded by 
NASA as an element of the Planet Finder Program.  JPL is managed for 
NASA by the California Institute of Technology.


\begin{thebibliography}{}

\bibitem[Augereau et al.(2001)]{auge01}Augereau, J.~C., Nelson, R.~P.,
 Lagrange, A.~M., Papaloizou, J.~C.~B., \& Mouillet, D. 2001, \aap, 370, 447

\bibitem[Banaszkiewicz et al.(1994)]{bana94}Banaszkiewicz, M., Fahr, H.~J., \&
Scherer, K. 1994, {\it Icarus}, 107, 358

\bibitem[Beauge \& Ferraz-Mello(1994)]{beau94}Beauge, C. \& Ferraz-Mello, S. 1994,
{\it Icarus}, 110, 239

\bibitem[Brouwer \& Clemence(1961)]{brou61}Brouwer, D. \& Clemence, G.~M. 1961,
Methods of Celestial Mechanics (New York: Academic Press)

\bibitem[Burns et al.(1979)]{burn79}Burns, J.~A., Lamy, P., \& Soter, S. 1979,
{\it Icarus}, 40, 1

\bibitem[Burrows et al.(1995)]{burr95}Burrows, C.~J., Krist, J.~E., Stapelfeldt, K.~R.,
and the WFPC2 Investigation Definition Team 1995, {\it BAAS}, 187, 3205


\bibitem[Dermott et al.(1994)]{derm94}Dermott, S.F., Jayaraman, S., Xu,  Y.~L., Gustafson, B.\AA.S. and Liou, J.-C. 1994, {\it Nature}, 369, 719

\bibitem[Dermott et al.(1985)]{derm85}Dermott, S.F., Nicholson, P.~D.,
Burns, J.~A., \& Houck, J.~R. 1985, in Properties and Interactions of
Interplanetary Dust, ed. R.~H. Giese \& P. Lamy (Dordrecht:Reidel), 395

\bibitem[Duncan et al.(1989)]{dunc89}Duncan, M., Quinn, T. \& Tremaine, S.
1989, {\it Icarus}, 82, 402

\bibitem[Fixsen \& Dwek(2002)]{fixs02}Fixsen, D.~J. \& Dwek, E. 2002, \apj, 578, 1009

\bibitem[Gold(1975)]{gold75}Gold, T. 1975, Icarus, 25, 489

\bibitem[Greaves et al.(1998)]{grea98}Greaves, J. S., Holland, W. S.,
Moriarty-Schieven, G., Jenness, T., Dent, W. R. F., Zuckerman, B.,
McCarthy, C., Webb, R. A., Butner, H. M., Gear, W. K. and Walker,
H. J. 1998, \apj, 506, L133.

\bibitem[Greenberg(1978)]{gree78}Greenberg, R. 1978, {\it Icarus}, 33, 62

\bibitem[Grun et al.(1985)]{grun85}Grun, E., Zook, H.~A., Fechtig, H. \& Giese,
R.~H. 1985, {\it Icarus}, 62, 244

\bibitem[Hauser et al.(1984)]{haus84}Hauser, M.~G., Gillett, F.~C., Low, F.~J.,
Gautier, T.~N., Beichman, C.~A., Aumann, H.~H., Neugebauer, G., Baud, B.,
Boggess, N., \& Emerson, J.~P. 1984, \apj, 278, L15
 


\bibitem[Holland et al.(2003)]{fomalhaut}Holland, W.~S., Greaves,
J.~S., Dent, W.~R.~F., et al. 2003, \apj, 582, 1141

\bibitem[Holland et al.(1998)]{holl98}Holland, W.~S., Greaves, J.~S.; Zuckerman, B.,
Webb, R.~A., McCarthy, C., Coulson, I.~M., Walther, D.~M., Dent, W.~R.~F., Gear, W.~K.,
Robson, I. 1998, Nature, 392, 788



\bibitem[Jackson \& Zook(1989)]{jack89}Jackson, A. A. and Zook, H. A. 1989, Nature, 337, 629

\bibitem[Koerner et al.(2001)]{koer01}Koerner, D~W., Sargent A.~I.,
\& Ostroff, N.~A., 2001, ApJ, 560, L181

\bibitem[Kuchner et al.(2000)]{kuch00}Kuchner, M.~J., Reach, W.~T.,
\& Brown, M.~E. 2000, {\it Icarus}, 145, 44

\bibitem[Kuchner \& Holman(2001)]{kuch01}Kuchner, M.~J. and Holman, M. 2001, presented at 
  the December 2001 Division of Planetary Sciences meeting

\bibitem[Lazzaro et al.(1994)]{lazz94}Lazzaro, D., Sicardy, B., Roques, F.
\& Greenberg, R. 1994, {\it Icarus}, 108, 59

\bibitem[Lecavelier Des Etangs et al.(1996)]{leca96}Lecavelier Des Etangs, A.,
Scholl, H., Roques, F., Sicardy, B., \& Vidal-Madjar, A. 1996, {\it Icarus}, 123,168

\bibitem[Liou \& Zook(1997)]{liou97}Liou, J.-C. and Zook, H.~A. 1997, {\it Icarus}, 128, 354

\bibitem[Liou \& Zook(1999)]{liou99}Liou, J.-C. and Zook, H.~A. 1999, \aj, 118, 580

\bibitem[Marcy \& Butler(2000)]{marc00}Marcy, G.~W. \& Butler, R.~P. 2000, \pasp, 768, 137 

\bibitem[Marzari \& Vanzani(1994)]{marz94}Marzari, F. \& Vanzani, V. 1994, \aap, 283, 275


\bibitem[Murray \& Dermott(1999)]{murr99}Murray, C.~D. \& Dermott, S.~F. 1999, 
Solar System Dynamics (New York: Cambridge Univ. Press)


\bibitem[Ozernoy et al.(2000)]{ozer00}Ozernoy, L.~M., Gorkavyi, N.~N.,
Mather, J.~C., \& Taidakova, T.~A. 2000, \apj, 537, L147

\bibitem[Quillen \& Thorndike(2002)]{quil02}Quillen, A.~C. \& Thorndike, S. 2002,
\apj, 578, L149

\bibitem[Reach(1991)]{reac91}Reach, W.~T. 1991, \apj, 369, 529

\bibitem[Reach et al.(1995)]{reac95}Reach, W.~T., Franz, B.~A., Weiland, J.~L.,
Hauser, M.~G., Kelsall, T.~N., Wright, E.~L., Rawley, G., Stemwedel, S.~W. and Spiesman, W.~J. 1995, Nature, 374, 521

\bibitem[Robertson(1937)]{robe37}Robertson, H.~P. 1937, \mnras, 97, 423

\bibitem[Roques et al.(1994)]{roqu94}Roques, F., Scholl, H.,
 Sicardy, B., Smith, B.~A. 1994, {\it Icarus}, 108, 37
 
\bibitem[Schneider et al.(1999)]{schn99} Schneider, G., Smith, B.~A., Becklin, E.~E., Koerner, D.~W., Meier, R., Hines, D.~C., Lowrance, P.~J., Terrile, R.~J., Thompson, R.~I., \& Rieke, M. 1999, \apj, 513, L127

\bibitem[Sicardy et al.(1993)]{sica93}Sicardy, B., Beauge, C., Ferraz-Mello, S.,
Lazzaro, D., \& Roques, F. 1993, {\it CeMDA}, 57, 373

\bibitem[Weidenschilling \& Jackson(1993)]{weid93}Weidenschilling,
S.~J. \& Jackson, A.~A. 1993, Icarus, 104, 244

\bibitem[Wisdom(1980)]{wisd80}Wisdom, J. 1980, \aj, 85, 1122

\bibitem[Wisdom(1983)]{wisd83}Wisdom, J. 1983, {\it Icarus},  56, 51

\bibitem[Wilner et al.(2002)]{wiln02}Wilner, D.~J., Holman, M.~J.,
Kuchner, M.~J., \& Ho, P.~T.~P. 2002, \apj, 569, L115

\bibitem[Wyatt \& Dent(2002)]{wyat02}Wyatt, M.~C. \& Dent, W.~R.~F. 2002, \mnras, 334, 589

\bibitem[Wyatt et al.(1999)]{wyat99}Wyatt, M.~C., Dermott, S.~F., Telesco, C.~M., Fisher, R.~S.,
 Grogan, K., Holmes, E.~K., \& Pi\~na, R.~K. 1999, \apj, 527, 918
 
\bibitem[Wyatt \& Whipple(1950)]{wyat50} Wyatt, S.~P. \& Whipple, F.~L.
1950, \apj, 111, 134

\end{thebibliography}
\end{document}